\documentclass[aps,prb,twocolumn,showpacs]{revtex4}
\usepackage{graphicx}
\usepackage{dcolumn}
\usepackage[bookmarks=false]{hyperref}
\hypersetup{pdfstartview=FitH}

\begin{document}

\title{Comment on "Estimate of the vibrational frequencies of spherical virus
particles"}

\author{Lucien Saviot}
\affiliation{Laboratoire de Recherche sur la R\'eactivit\'e des Solides,
UMR 5613 CNRS - Universit\'e de Bourgogne\\
9 avenue A. Savary, BP 47870 - 21078 Dijon - France}
\email{lucien.saviot@u-bourgogne.fr}
\author{Daniel B. Murray}
\affiliation{Department of Physics,
Okanagan University College,
Kelowna, British Columbia, Canada V1V 1V7}
\email{dmurray@ouc.bc.ca}
\author{Alain Mermet}
\email{mermet@pcml.univ-lyon1.fr}
\affiliation{Laboratoire de Physico-Chimie des Mat\'eriaux Luminescents, UMR 5620
CNRS - Universit\'e Lyon I\\
43, boulevard du 11 Novembre 69622 Villeurbanne Cedex - France}
\author{Eug\`ene Duval}
\email{duval@pcml.univ-lyon1.fr}
\affiliation{Laboratoire de Physico-Chimie des Mat\'eriaux Luminescents, UMR 5620
CNRS - Universit\'e Lyon I\\
43, boulevard du 11 Novembre 69622 Villeurbanne Cedex - France}
\date{\today}
\begin{abstract}
This comment corrects some errors which appeared in the calculation of an
elastic sphere eigenenergies. As a result, the symmetry of the mode having
the lowest frequency is changed.
Also a direction for calculating the damping of these modes for embedded
elastic spheres is given.
\end{abstract}
\maketitle
In a recent article\cite{FordPRE03}, L. H. Ford discusses the normal modes of
vibration of spherical virus particles using a liquid drop model and an
elastic sphere model.

However the analysis errs in calculating the energy of
the $n=0$ spheroidal mode with the elastic sphere model.
This same error is recurrent in the
literature \cite{eringen,Fujii91,Tanaka93,Kuok03}.  Some authors
have since corrected them\cite{erratumFujii91}.  The original
Lamb paper\cite{lamb1882} was correct and an explanation
is available elsewhere\cite{saviot96}.
This error results in additional energies which do not correspond
to any vibration eigenmode.
It gives a wrong energy for the lowest $n=0$ spheroidal
mode comparable to the one of the lowest $n=2$ spheroidal mode.
For materials with positive Poisson ratio, it is impossible to have
the energy for the fundamental $n=0$ mode smaller than the
energy for the fundamental $n=2$ one.

Equation 3 in the paper is valid for $n$
different from zero. For $n$ equal to zero, $S_{\rm rp}(0,x,y) = 0$
should be used
instead (see equation 4). Moreover, expressions given in equations 4 and 7 are
invalid: there is no square dependence and the sign of $S_{\rm ts}$ is wrong.
With correct calculations, the lowest frequencies for the elastic sphere model
are reached with $n=2$ modes and frequencies are a bit lower (9.2, 9.4 and 4.6
GHz for Nylon, Polystyrene and Polyethylene respectively).

We also would like to point out that these normal modes are damped when the
virus particle is embedded inside a liquid. Estimations of
damping can be made using the "complex frequency" approach described
elsewhere\cite{dubrovskiy81,saviotPRB03}. When the virus is inside water for
example, there is not much acoustic impedance mismatch at the surface.
For this reason the normal modes will be broad and have short lifetimes.
Therefore, the objective of killing viruses by sending out
sound waves that resonate and destroy them is probably
unworkable in such a configuration.

\bibliographystyle{utphys}
\bibliography{biblio}
\end{document}